\documentclass[runningheads,a4paper]{llncs}

\usepackage{amssymb}
\usepackage{graphicx}

\usepackage{url}
\urldef{\mailsa}\path|{alfred.hofmann, ursula.barth, ingrid.haas, frank.holzwarth,|
\urldef{\mailsb}\path|anna.kramer, leonie.kunz, christine.reiss, nicole.sator,|
\urldef{\mailsc}\path|erika.siebert-cole, peter.strasser, lncs}@springer.com|    
\newcommand{\keywords}[1]{\par\addvspace\baselineskip
\noindent\keywordname\enspace\ignorespaces#1}

\begin{document}

\mainmatter  

\title{Semi-Automated Virtual Unfolded View Generation Method of Stomach \\from CT Volumes}

\titlerunning{Semi-Automated Virtual Unfolded View Generation Method}

%
%
\author{Masahiro Oda \inst{1}
\and
Tomoaki Suito \inst{1}
\and
Yuichiro Hayashi \inst{2}
\and
Takayuki Kitasaka \inst{3}
\and\\
Kazuhiro Furukawa \inst{4}
\and
Ryoji Miyahara \inst{4}
\and
Yoshiki Hirooka \inst{5}
\and
Hidemi Goto \inst{4}
\and\\
Gen Iinuma \inst{6}
\and
Kazunari Misawa \inst{7}
\and
Shigeru Nawano \inst{8}
\and
Kensaku Mori \inst {2,1}}
\authorrunning{Masahiro Oda et al.}

\institute{Graduate School of Information Science, Nagoya University,\\
\and
Information and Communications Headquarters, Nagoya University,\\
\and 
School of Information Science, Aichi Institute of Technology,\\
\and
Graduate School of Medicine, Nagoya University,\\
\and
Department of Endoscopy, Nagoya University Hospital,\\
\and 
National Cancer Center,\\
\and 
Aichi Cancer Center,\\
\and 
International University of Health and Welfare Mita Hospital
}

%
%

\toctitle{Semi-Automated Virtual Unfolded View Generation Method of Stomach from CT Volumes}
\tocauthor{Masahiro Oda}
\maketitle

\begin{abstract}
CT image-based diagnosis of the stomach is developed as a new way of diagnostic method.
A virtual unfolded (VU) view is suitable for displaying its wall.
In this paper, we propose a semi-automated method for generating VU views of the stomach.
Our method requires minimum manual operations.
The determination of the unfolding forces and the termination of the unfolding process are automated.
The unfolded shape of the stomach is estimated based on its radius.
The unfolding forces are determined so that the stomach wall is deformed to the expected shape.
The iterative deformation process is terminated if the difference of the shapes between the deformed shape and expected shape is small.
Our experiments using 67 CT volumes showed that our proposed method can generate good VU views for 76.1\% cases.
\end{abstract}

\section{Introduction}

In Japan, the mortality rate of stomach cancer is the second highest among cancer-related mortality \cite{Isobe11}.
Treatment of stomach cancer in the early stages is crucial.
Gastric roentgenography and gastrofiberscopy are currently performed as the diagnostic methods of stomach cancer.
But these methods are physically and mentally painful for patients.
In recent years, a CT image-based diagnostic method of the stomach has been developed as an alternative choice \cite{Furukawa11} that utilizes virtual gastroscopic views generated from CT images.
Although CT image-based stomach diagnosis systems greatly reduce the inspection time for patients, 
physicians need to manually change the viewpoints and the viewing directions of the virtual gastroscopic views many times during diagnosis.
To reduce this load of physicians, a virtual unfolded (VU) view of the stomach is suitable for displaying the stomach wall.
This view enables physicians to observe the stomach wall at a glance.

Much research has been reported on the VU view generation of hollow organs.
Most generate VU views of the colon \cite{Wang98,Bartroli01}.
Because these methods generate views in real-time, their unfolding processes do not follow the physical properties of organ deformation.
VU view generation methods of the stomach have been proposed based on surface model \cite{Mori01} and volumetric model \cite{Mori05}.
Although these methods simulate realistic deformations, they require the following complicated manual operations: (1) incision line determination, (2) unfolding force determination, and (3) the termination of unfolding processes.
Truong et al. \cite{Truong08} automated (2) the unfolding force determination process.
However, the results of their method heavily depend on the results of incision line determination.
Their method requires many trial-and-error corrections in incision line determination.
Suito et al. \cite{Suito12} presented an automated method of (1) the incision line determination process.
These researches only automate one of the manual processes.
Therefore, these VU view generation methods of the stomach still require complicated manual operations.
They are time-consuming and the quality of their generated VU views heavily depends on the skill of the user.

In this paper, we propose a semi-automated method for generating VU views of the stomach from CT volumes.
The contribution of this paper is first presentation of semi-automated method that can drastically reduce manual operations in generation of VU views of the stomach.
Manual operation required in our method is just specifying two positions on CT images: the cardia and the pylorus.
All other processes are automated in our method.
We determine the incision line using a previous method \cite{Suito12}.
Then a stomach wall model is generated from a stomach wall region extracted from a CT volume.
Unfolding forces are added to the model.
We newly introduce a method that automatically calculates the direction of the unfolding forces for this.
The expected unfolded shape is estimated based on its diameter.
Unfolding force is determined so that the stomach wall is deformed to the expected shape.
Newmark-$\beta$ \cite{Newmark59} method is introduced to simulate elastic deformation of the model.
Because the unfolding process is performed iteratively, it must be terminated after appropriate iterations.
For the termination of the unfolding process, we define a criterion that evaluates the progress of the unfolding process based on the differences of the shapes between the deformed stomach shape and the expected unfolded shape of the stomach.

\section{Method}

\subsection{Preprocessing}

In the preprocessing step, we extract an air region in the stomach, a stomach wall region, and the centerline of the stomach region from abdominal CT volumes \cite{Suito12}.
Then the incision lines are determined using the methods shown in \cite{Suito12}.
Since the incision line determination process \cite{Suito12} requires the positions of the cardia and the pylorus, these points are manually specified on CT volumes by mouse-click.

\subsection{Definition of Stomach Wall Model}

We utilize a previously proposed stomach wall model \cite{Truong08} to simulate the deformation.
The model consists of a set of hexahedra covering the stomach wall region.
The center of each hexahedron is a voxel in the stomach wall region.
The length of each hexahedron edge in the $x, y$, and $z$ directions are $d, d$, and $\hat{d}$ voxels, respectively.
Here, $\hat{d} = d \cdot (\rm{pixel\ spacing})/(\rm{slice\ spacing})$.
The hexahedra on the incision line are removed from the stomach wall model.
Each hexahedron is converted into elastic model by placing mass-points, springs, and dampers on vertices, edges, and diagonal lines \cite{Truong08}.

\begin{figure}[tb]
 \begin{minipage}{1.0\linewidth}
  \begin{center}
   \includegraphics[width=105mm]{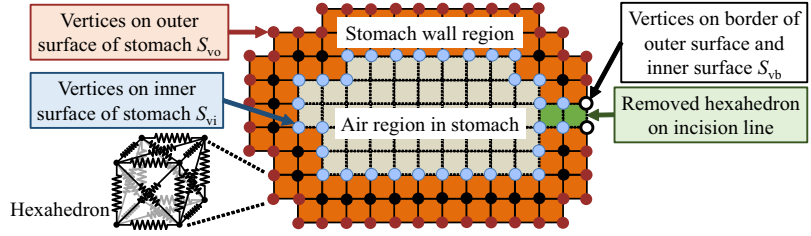}
  \end{center}
 \end{minipage}
\caption{Sets of vertices of hexahedra $S_{\rm vo}$, $S_{\rm vi}$, and $S_{\rm vb}$ on stomach wall model.}
\label{fig:vertices}
\end{figure}

We define three sets of the hexahedra vertices in the stomach wall model: $S_{\rm vo}$, $S_{\rm vi}$, and $S_{\rm vb}$.
A set of vertices on the outer and inner surfaces of the stomach wall model is represented as $S_{\rm vo}$ and $S_{\rm vi}$, respectively.
The inner surface of the stomach wall model is a set of the faces of hexahedra in contact with the air region in the stomach and the incision line.
A set of vertices near the incision line, which is included in both $S_{\rm vo}$ and $S_{\rm vi}$, are defined as $S_{\rm vb}$ (Fig. \ref{fig:vertices}).

\subsection{Determination of Unfolding Force}

The unfolding forces are added to the vertices of hexahedra on the incision line, which is unfolded to a planar shape by the forces.
We first define an {\it unfolded plane} on which the stomach wall model is stretched.
Then we obtain the {\it destination points} on the plane, where the vertices on the incision line must reach after unfolding.

\subsubsection{Determination of Unfolded Plane}

\begin{figure}[tb]
 \begin{minipage}[b]{0.32\linewidth}
  \begin{center}
   \includegraphics[width=39mm]{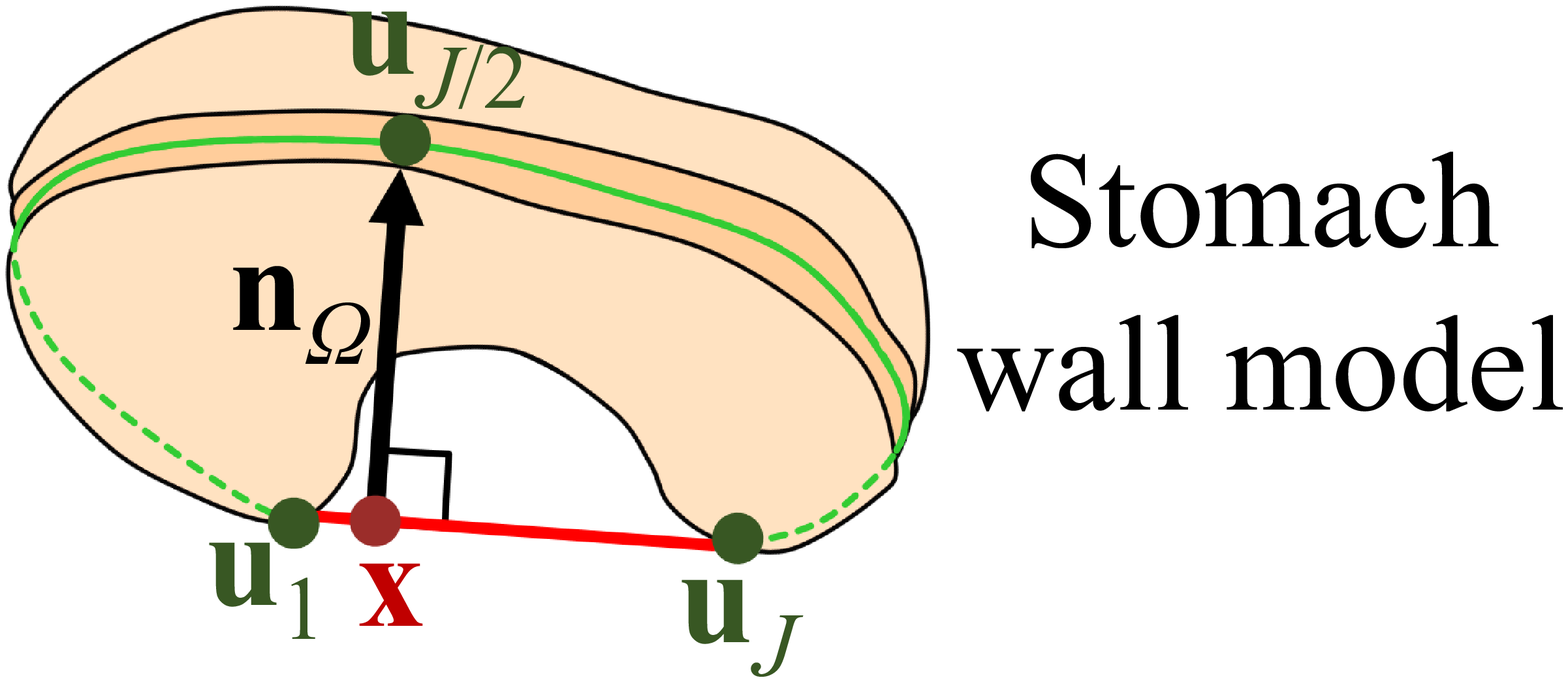}\\
   (a)
  \end{center}
 \end{minipage}
 \begin{minipage}[b]{0.32\linewidth}
  \begin{center}
   \includegraphics[width=35mm]{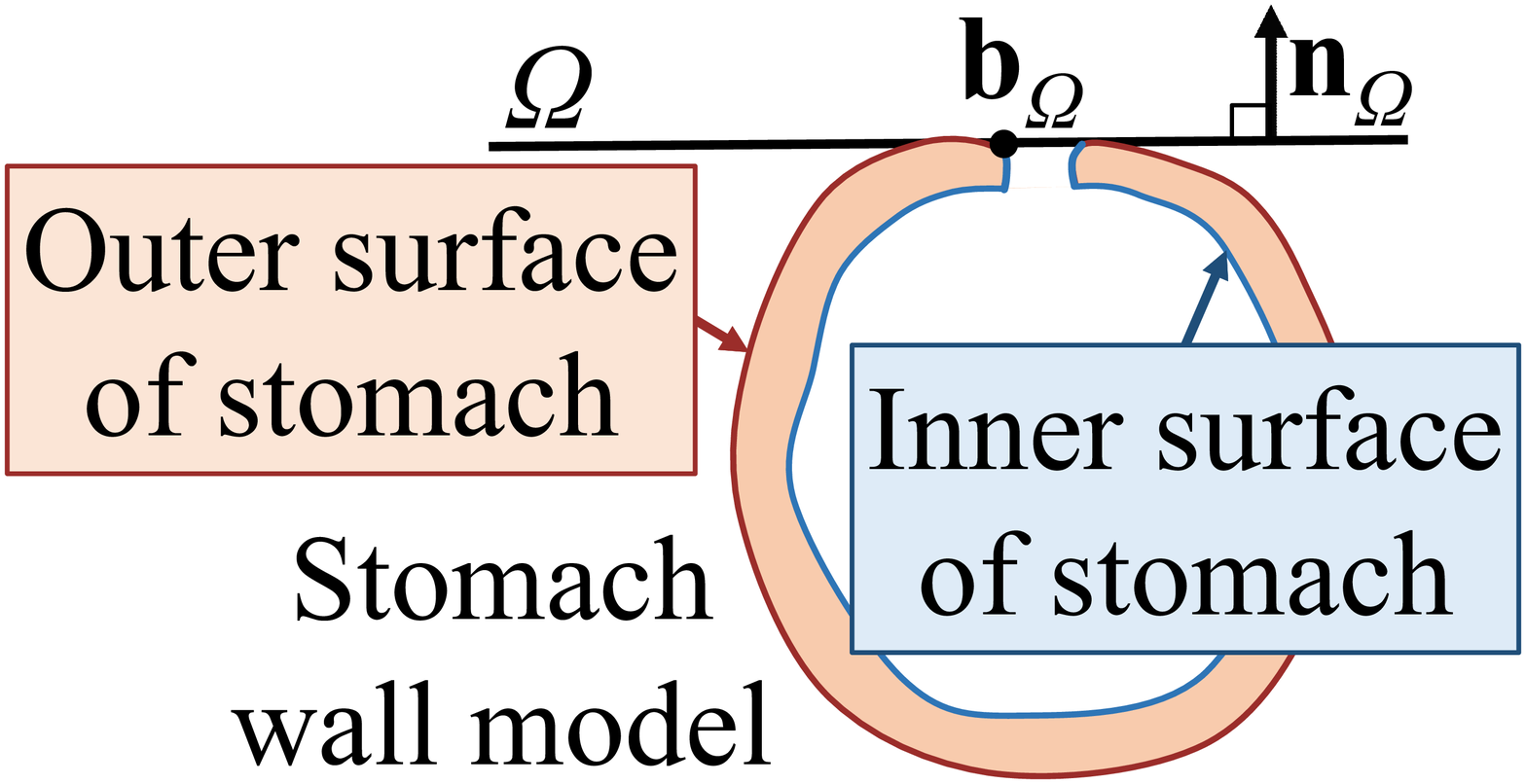}\\
   (b)
  \end{center}
 \end{minipage}
 \begin{minipage}[b]{0.32\linewidth}
  \begin{center}
   \includegraphics[width=30mm]{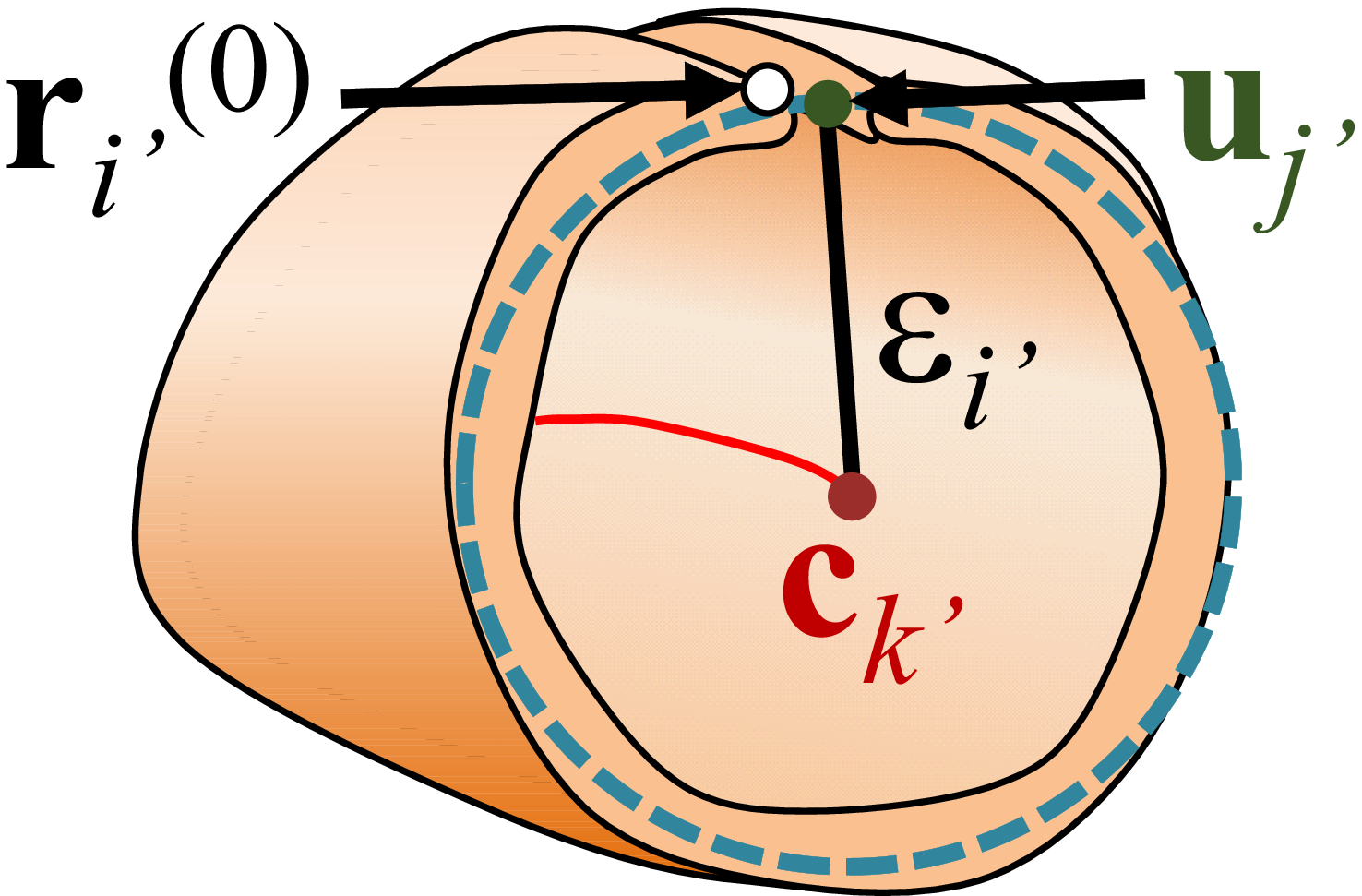}\\
   (c)
  \end{center}
 \end{minipage}
\caption{(a) Normal vector ${\bf n}_{\Omega}$, (b) point ${\bf b}_{\Omega}$ on $\Omega$, and (c) radius of stomach $\epsilon_{i'}$ at $V_{i'}$.}
\label{fig:omega}
\end{figure}

Unfolded plane $\Omega$ is uniquely defined by its normal vector ${\bf n}_{\Omega}$ and point ${\bf b}_{\Omega}$ on it as
\begin{eqnarray}
{\bf n}_{\Omega} & = & ({\bf u}_{J/2} - {\bf x}) / \parallel {\bf u}_{J/2} - {\bf x} \parallel,\\
{\bf b}_{\Omega} & = & {\bf r}^{(0)}_{m}, \ m = {\rm argmax}_{V_{i} \in S_{\rm vi}} \left( {\bf r}^{(0)}_{i} \cdot {\bf n}_{\Omega} \right),
\end{eqnarray}
where ${\bf u}_{j} \ (j=1,\ldots,J)$ is a point sequence forming the incision line.
${\bf x}$ is a point on a line segment, which connects ${\bf u}_{1}$ and ${\bf u}_{J}$, and satisfies $({\bf x} - {\bf u}_1)\cdot({\bf u}_{J/2} - {\bf x}) = 0$.
$V_{i} \ (i=1,\ldots,I)$ is the $i$-th vertex in the stomach wall model.
${\bf r}^{(\alpha)}_{i}$ is the position of $V_{i}$ after $\alpha$ iterations of the unfolding process.
The positions of ${\bf n}_{\Omega}$ and ${\bf b}_{\Omega}$ are shown in Fig. \ref{fig:omega}(a) and (b).

\subsubsection{Determination of Destination Point}

This process consists of three steps: (Step 1) Calculation of stomach radius, (Step 2) Calculation of base line, and (Step 3) Determination of destination point.
The thick and thin parts of the stomach wall must be unfolded widely and narrowly.
We compute the width of the unfolded view based on the radius of the stomach at each point on the incision line.
Step 1 calculates the radius of the stomach.
Step 2 computes the base line around which the stomach is unfolded.
Step 3 determines the points where each vertex of the incision line is forced to be moved.

\noindent {\bf (Step 1) Calculation of Stomach Radius}: 
We obtain the radius of the stomach for each vertex existing on the incision line.
The point sequence on the centerline of the stomach is described as ${\bf c}_{k} \ (k=1,\ldots,K)$.
${\bf c}_{k}$ that is closest to ${\bf u}_{j}$ is obtained as ${\bf c}_{k'}$ whose index is calculated by
\begin{equation}
k' = {\rm argmin}_{1 \leq k \leq K} \parallel {\bf c}_{k}-{\bf u}_{j} \parallel.
\end{equation}
$V_{i'} \in S_{\rm vb} \ (i'=1,\ldots,I')$ is the $i'$-th vertex existing on the incision line on the stomach wall model.
The index of ${\bf u}_{j}$, which is closest to ${\bf r}^{(0)}_{i'}$ is obtained by
\begin{equation}
j' = {\rm argmin}_{1 \leq j \leq J} \parallel {\bf r}^{(0)}_{i'}-{\bf u}_{j} \parallel.
\end{equation}
The radius of the stomach (Fig. \ref{fig:omega}(c)) at $V_{i'}$ is calculated by
\begin{equation}
\epsilon_{i'} = \parallel {\bf c}_{k'} - {\bf u}_{j'} \parallel.
\end{equation}

\begin{figure}[tb]
 \begin{minipage}[b]{0.48\linewidth}
  \begin{center}
   \includegraphics[width=35mm]{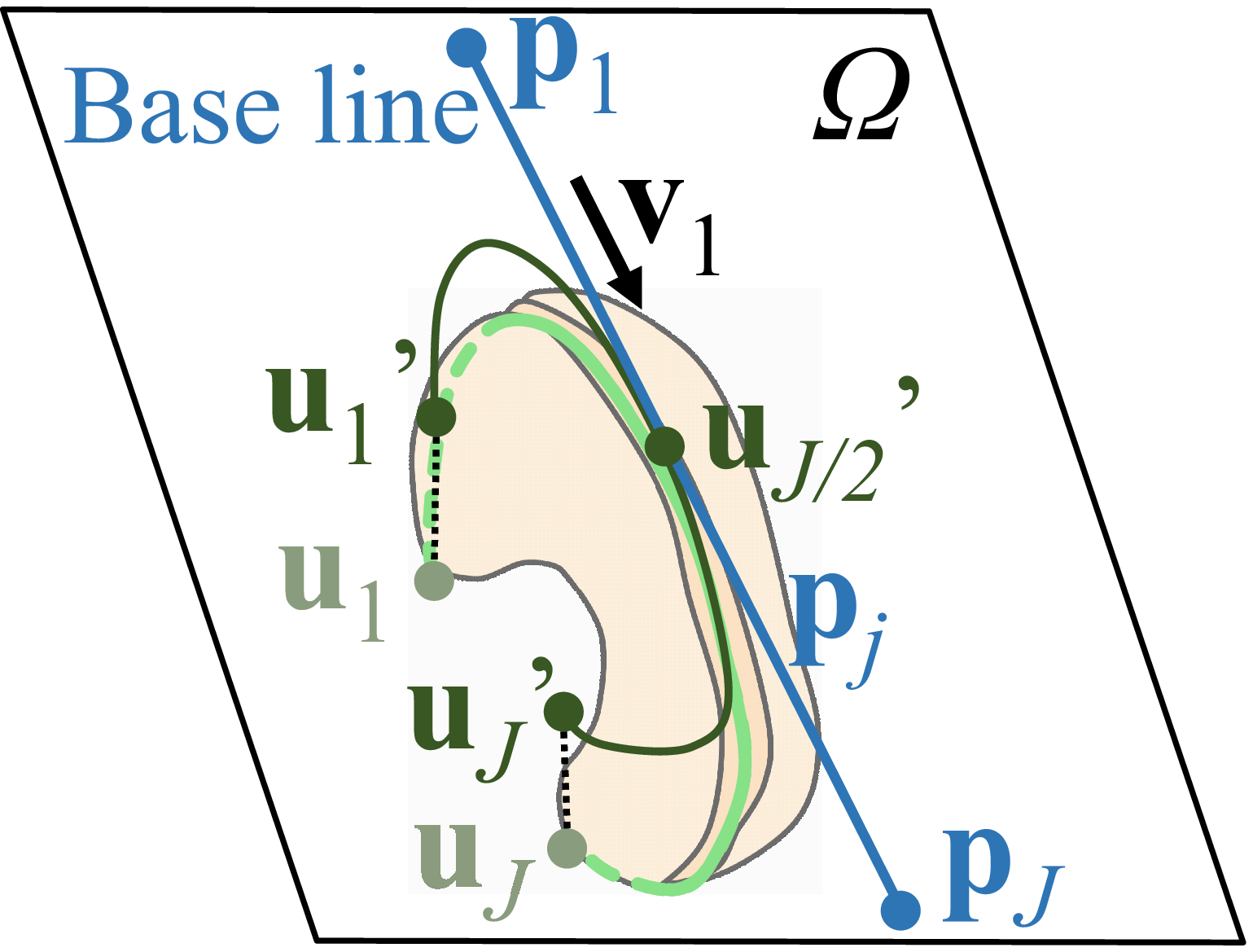}\\
   (a)
  \end{center}
 \end{minipage}
 \begin{minipage}[b]{0.48\linewidth}
  \begin{center}
   \includegraphics[width=35mm]{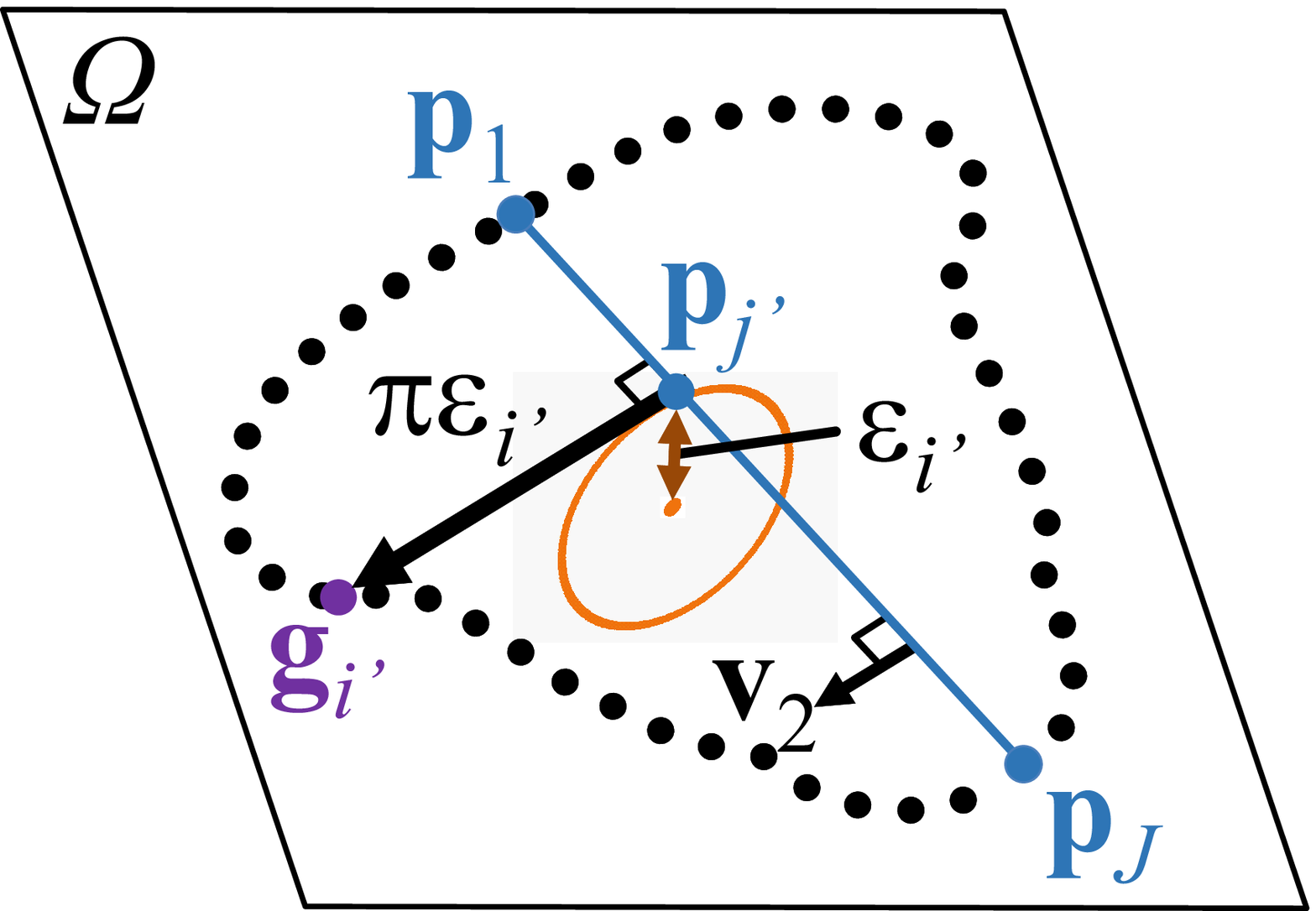}\\
   (b)
  \end{center}
 \end{minipage}
\caption{(a) Base line ${\bf p}_{j}$ calculated from projected incision line ${\bf u}'_{j}$. (b) Destination point ${\bf g}_{i'}$ corresponds to $V_{i'}$.}
\label{fig:destinationpoint}
\end{figure}

\noindent {\bf (Step 2) Calculation of Base Line}: 
The base line is a long axis of the point set on the incision line projected on the unfolded plane.
Each point ${\bf u}_{j}$ is projected onto the closest point ${\bf u}'_{j}$ on the unfolded plane.
We apply principal component analysis to ${\bf u}'_{j}$ to obtain the first and second eigenvectors of ${\bf u}'_{j}$.
The first and second eigenvectors are ${\bf v}'_{1}$ and ${\bf v}'_{2}$.
The direction of the base line is given by
\begin{eqnarray}
{\bf v}_1 = \left\{ 
	\begin{array}{ll}
	{\bf v}'_1 / \parallel {\bf v}'_1 \parallel,	&	\ if \ {\bf v}'_1 \cdot ({\bf u}'_J - {\bf u}'_1) < 0,\\
	-{\bf v}'_1 / \parallel {\bf v}'_1 \parallel,	&	\ otherwise.
	\end{array}
\right. 
\end{eqnarray}
The base line is represented as a set of points and defined by
\begin{eqnarray}
{\bf p}_{j}	= \left\{
	\begin{array}{ll}
	{\bf u}'_{J/2} - {\bf v}_1\sum_{z=j}^{J/2-1} \parallel {\bf u}'_{z+1} - {\bf u}'_z \parallel,	&	\ if \ j=1,\ldots,J/2-1, \\
	{\bf u}'_{J/2},																					&	\ if \ j=J/2,\\
	{\bf u}'_{J/2} + {\bf v}_1\sum_{z=J/2}^{j-1} \parallel {\bf u}'_{z+1} - {\bf u}'_z \parallel,	&	\ if \ j=J/2+1,\ldots,J.
	\end{array}
\right.
\end{eqnarray}
The position of ${\bf p}_{j}$ is shown in Fig. \ref{fig:destinationpoint}(a).

\noindent {\bf (Step 3) Determination of Destination Point}: 
Destination points ${\bf g}_{i'}$ are determined based on the base line and the radius of the stomach.
${\bf g}_{i'}$ is positioned $\pi \epsilon_{i'}$[mm] away from the base line on the unfolded plane (Fig. \ref{fig:destinationpoint}(b)):
\begin{eqnarray}
{\bf g}_{i'} = \left\{
	\begin{array}{ll}
	{\bf p}_{j'} + \pi\epsilon_{i'}{\bf v}_2,	&	\ if \ \{({\bf u}_{j'} - {\bf c}_{k'}) \times ({\bf r}^{(0)}_{i'} - {\bf c}_{k'})\} \cdot ({\bf u}_{j'} - {\bf u}_{j'-1}) \geq 0,\\
	{\bf p}_{j'} - \pi\epsilon_{i'}{\bf v}_2,	&	\ otherwise,
	\end{array}
\right.
\end{eqnarray}
${\bf v}_{2}$ is unit vector perpendicular to the base line obtained by
\begin{eqnarray}
{\bf v}_2 = \left\{ 
	\begin{array}{ll}
	{\bf v}'_2 / \parallel {\bf v}'_2 \parallel,	&	\ if \ ({\bf v}'_2 \times {\bf v}_1) \cdot {\bf n}_\Omega < 0,\\
	-{\bf v}'_2 / \parallel {\bf v}'_2 \parallel,	&	\ otherwise.
	\end{array}
\right. 
\end{eqnarray}

\subsubsection{Unfolding Force Determination} \label{sssec:ufd}

The direction of the unfolding force for $V_{i'}$ calculated as
\begin{eqnarray}
{\bf e}^{(\alpha)}_{i'} = {\bf g}_{i'} - {\bf r}^{(\alpha)}_{i'},
\end{eqnarray}
where $\alpha$ is the number of iterations in the iterative unfolding process.
${\bf e}^{(\alpha)}_{i'}$ is updated at each iteration step in the unfolding process.

\subsection{Unfolding Process Based on Elastic Deformation}

We deform the stomach wall model by Newmark-$\beta$ method \cite{Newmark59} with the forces computed above.
The shell-like volumetric model between the inner and outer surfaces shown in Fig. \ref{fig:vertices} is deformed here.
We employ the elastic deformation procedure shown in \cite{Truong08}.
This method \cite{Truong08} manually terminated the iterative unfolding process.
We propose a criterion to evaluate progress of the unfolding process in order to determine termination of deformation.

We introduce new metric that measures the average distance between ${\bf r}^{(\alpha)}_{i'}$ and its destination point.
The metric is described as
\begin{eqnarray}
D^{(\alpha)} = \frac{1}{N} \sum_{V_{i'} \in S_{\rm vb}} \parallel {\bf r}^{(\alpha)}_{i'} - {\bf g}_{i'} \parallel,
\end{eqnarray}
where $N$ is the number of vertices included in $S_{\rm vb}$.
If the change of the average distance is small in the iteration of the unfolding process, the stomach is adequately stretched.
The iterative unfolding process is terminated if condition
\begin{equation}
\parallel D^{(\alpha-1)}-D^{(\alpha)} \parallel \leq \kappa
\end{equation}
is satisfied.

\subsection{Unfolded View Generation}

After the stomach wall model deformation, we generate an unfolded volume by using relationship between the stomach wall models of pre- and post-deformations.
The unfolded view is generated by volume-rendering of the unfolded volume.

\section{Experiments and Results}

\begin{figure}[tb]
 \begin{minipage}{1.0\linewidth}
  \begin{center}
   \includegraphics[width=95mm]{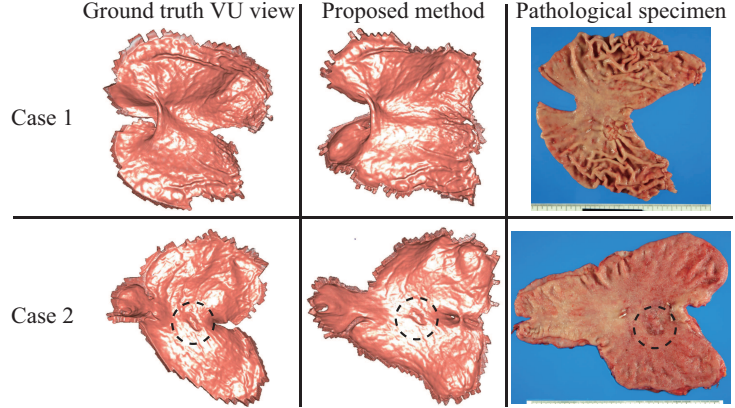}	
  \end{center}
 \end{minipage}
\caption{Ground truth VU views, VU views generated by proposed method, and flattened pathological specimens of two cases. Circles indicate positions of a cancer.}
\label{fig:result}
\end{figure}

We evaluated the proposed method by using 67 CT volumes that were taken from the distended state of the stomach using a foaming agent at two hospitals.
The followings are the acquisition parameters of the CT volumes: image size: 512$\times$512 pixels, number of slices: 371-644, pixel spacing: 0.625-0.774 mm, slice spacing: 0.4-1.0 mm, slice thickness: 0.5-1.0 mm.
Parameter values $d$ = 8 voxels and $\kappa = 0.5$ were experimentally chosen.
The quality of the VU views was visually evaluated by surgeons and engineering researchers by giving {\it excellent} (All region on VU view is visible from a view point), {\it good} (One small failure part (overlapping, bending, or broken) exists on VU view), {\it fair}, and {\it bad} (Area of VU view where there are not visible from a view point by influence of failure parts is over 30\% and 50\%, respectively) classifications.
The generated VU views were compared with the flattened pathological specimens of the resected stomachs from the patients.
We prepared ground truth VU views with manual input of the incision lines and the unfolding forces.
The generated VU views are shown in Fig. \ref{fig:result}.
The numbers of cases rated as top one ({\it excellent}) and two ({\it excellent} or {\it good}) classifications were 26 (38.8\%) and 51 (76.1\%) cases.

Since the quality of the VU views depends on the value of $\kappa$, we generated VU views by changing $\kappa$ values for 19 cases.
Table \ref{table:kappa} shows the numbers of good results and cases that caused overlapping, bending, and broken parts of the stomach wall in the VU views.

We compared the computation time in VU view generation of the proposed and a previous \cite{Mori05} methods (Computer: two Intel Xeon 3.33GHz processors, 4GB RAM).
It took about 19 seconds for the manual input 
and 458 seconds for the automated process.
In the previous method \cite{Mori05}, manual processes including the incision line determination, the unfolding force determination, and the termination of the unfolding processes took about 900 seconds.

\begin{table}[tb]
 \begin{minipage}{1.0\linewidth}
  \begin{center}
   \caption{Numbers of good results and numbers of cases that caused overlapping, bending, and broken parts of stomach wall on generated VU views for various $\kappa$.}
   \begin{tabular}{c|c|c|c|c|c|c|c|c|c|c} \hline
	$\kappa$	&	0.1	&	0.2	&	0.3	&	0.4	&	0.5	&	0.6	&	0.7	&	0.8	&	0.9	&	1.0	\\ \hline \hline
	Good		&	7	&	10	&	10	&	10	&	10	&	9	&	8	&	7	&	7	&	7	\\ \hline
	Overlapping	&	5	&	4	&	2	&	2	&	2	&	2	&	1	&	0	&	0	&	0	\\ \hline
	Bending		&	10	&	7	&	8	&	7	&	6	&	7	&	7	&	7	&	7	&	7	\\ \hline
	Broken		&	1	&	1	&	2	&	2	&	2	&	2	&	3	&	4	&	4	&	4	\\ \hline
   \end{tabular}
   \label{table:kappa}
  \end{center}
 \end{minipage}
\end{table}

\section{Discussion}

From Fig. \ref{fig:result}, the VU views generated by our proposed method clearly represent the shapes of a cancer and the stomach wall.
Although the shapes of the views generated by the proposed method slightly differ from the ground truth VU views, such surface shapes as 
the cancer on the stomach wall were well observed.
The surface shape is important for finding lesions.
These results are applicable to other data judged as "Good".
Therefore, the VU views generated using our proposed method are applicable for diagnosis.

From Table \ref{table:kappa}, the numbers of broken parts in the VU views become larger when large values of $\kappa$ are used.
When $\kappa$ is large, the iteration of the unfolding process is terminated even if the stomach wall model is still largely being deformed.
In such cases, several parts of the stomach wall model are greatly stretched.
Largely stretched parts of the stomach wall model may cause broken parts in the VU views.
The numbers of the overlapping parts on the VU view becomes larger for smaller $\kappa$.
In this case, the iteration number of the unfolding process becomes large because the iteration continues until the incision lines of the model reach the destination points.
In the unfolding process \cite{Truong08}, forces are added to the vertices on the outer and inner surfaces of the stomach wall models.
The magnitude of these forces increase as iteration progresses.
This cause overlapping of the VU views.
To minimize broken or overlapping parts, $\kappa = 0.5$ was selected as an optimal value for testing 67 cases.

This paper presented a semi-automated VU view generation method of the stomach.
The automation of VU view generation also achieves qualitative stable VU views.
Quality of VU views of the previous method heavily depends on experience levels of users.
Our method generates VU views with minimal influence of user experience level.
It also has potential to explore new diagnostic method of the stomach like as screening examinations of early gastric cancers or planning of gastric cancer surgery.

\section{Conclusion}

This paper presented a semi-automated VU view generation method.
The determination of the unfolding forces and the termination of the unfolding were automated in our method.
Experiments using 67 CT volumes showed that our proposed method can generate VU views with same quality as manually generated ones.
Future work includes the automation of cardia and pylorus detection, unfolding of twisted stomachs, and evaluation using more samples.

\subsubsection*{Acknowledgments.}
Parts of this research were supported by the MEXT, the JSPS KAKENHI Grant Numbers 21103006, 25242047, and the Kayamori Foundation of Informational Science Advancement.

\end{document}